\documentclass[12pt]{article}

\usepackage{amsfonts}
\usepackage{amsmath}
\usepackage{amsthm}
\usepackage[onehalfspacing]{setspace}
\usepackage{natbib}
\usepackage{amssymb}
\usepackage{latexsym}
\usepackage{appendix}
\usepackage{float}
\usepackage{graphicx}
\usepackage[inline]{enumitem}
\usepackage{chngcntr}
\usepackage[margin=1in]{geometry}
\usepackage[colorlinks]{hyperref}

\usepackage{framed}
\usepackage{nicefrac}
\usepackage{bm}
\usepackage{bbm}
\usepackage{url}

\usepackage{multirow,array}
\usepackage{comment}

\hypersetup{citecolor=blue,linkcolor=blue}

\newcommand{\supp}{\mathrm{Supp}}
\newcommand{\grand}{\hat{\Sigma}}
\newcommand{\spdom}{\tilde{\Sigma}}

\newcommand{\dom}{\Gamma}

\newtheorem{lemma}{Lemma}
\newtheorem{theorem}{Theorem}
\newtheorem{corollary}{Corollary}

\newtheorem{definition}{Definition}

\newtheorem{proposition}{Proposition}
\newtheorem{remark}{Remark}[section]

\newcommand{\citepos}[1]{\citeauthor{#1}'s \citeyearpar{#1}}

\makeatletter
\renewenvironment{proof}[1][\proofname] {\par\pushQED{\qed}\normalfont\topsep6\p@\@plus6\p@\relax\trivlist\item[\hskip\labelsep\bfseries#1\@addpunct{.}]\ignorespaces}{\popQED\endtrivlist\@endpefalse}
\makeatother

\title{Iterated Elimination of Dominated Mixed Strategies}
\author{Federico Echenique and Amanda Friedenberg\thanks{The first author came across the question of comparing the number of rounds in the pure and mixed IEDS procedures while teaching first-year graduate game theory. The second author came up with the answer that is presented here.}}

\begin{document}

\maketitle

\begin{abstract}
This note compares two iterated dominance procedures, one based on eliminating mixed strategies and one based on eliminating pure strategies. It shows that, in compact and continuous games, the mixed-strategy procedure can trail the pure-strategy procedure by at most one round: If the pure-strategy procedure terminates at round $M$, the mixed-strategy procedure terminates at either $M$ or $M+1$, and this bound is tight. It also provides two characterizations of the mixed-strategy procedure that are round-for-round equivalent to it and that require checking for improvements only against opponents' pure strategies.
\end{abstract}

\section{Introduction}

Iterated dominance is a foundational solution concept in game theory. In finite games, it is equivalent to \citet{bernheim1984rsb} and \citepos{pearce1984rsb} concept of rationalizability.  Thus, a strategy survives $m$-rounds of iterated dominance if and only if it is consistent with rationality and $(m-1)^{\text{th}}$-order belief of rationality.  (See, e.g., \citealp{tan1988bayesian}, \citealp{brandenburger1987rac} for classic statements of the result; see \citealp{battigalli2002sba}, \citealp{friedenberg-iterated} for modern statements.)

In applying iterated dominance (and rationalizability), there is often an implicit assumption: that players choose pure strategies. This note instead studies iterated dominance when players explicitly choose mixed strategies. This is desirable for two reasons. First, it is a model of choice behavior consistent with mixed-strategy Nash equilibrium; as such, it lets us consider rationalizability and equilibrium under a single unified framework. Second, there is substantial empirical support for the idea that players' choices are stochastic; that is, allowing for mixed strategies is not just a theoretical convenience but a plausible behavioral assumption.

Interestingly, a mixed strategy can be dominated even if each pure strategy in its support is undominated. At least in principle, this can complicate a researcher's ability to infer players' reasoning from their observed behavior. If the researcher only observes pure strategies---and not whether players chose pure vs.\ mixed strategies---the researcher may not know if they have observed an undominated choice of a pure strategy, or the realization of a potentially dominated mixed strategy. This inference problem would be particularly stark if there is a large disconnect between the number of rounds required to eliminate pure strategies vs.\ the number of rounds required to eliminate mixed strategies. 

This motivates the question: How far can the mixed-strategy elimination procedure trail the pure-strategy procedure? The question is particularly sharp in finite games. There, the dominance procedure based on pure strategies will terminate at some finite round $M$. But, the mixed-strategy extension of the game is an infinite game. Does the procedure based on mixed strategies terminate at some finite round? If so, at what round?

We establish an upper bound on the divergence between iterated dominance based on pure strategies and iterated dominance based on mixed strategies: In any compact and continuous game, the procedure based on mixed strategies can trail the procedure based on pure strategies by at most one step. That is, if the procedure based on pure strategies terminates at round $M$, the procedure based on mixed strategies will either terminate at $M$ or $M+1$. (See Theorem \ref{thm:main}.) The bound is tight. 

To establish the result, we introduce an intermediate concept, termed grandfather dominance. Grandfather dominance is a procedure on mixed strategies that asks whether a mixture is dominated relative to the ``grandfather'' set of pure strategies---i.e., relative to a set of pure strategies that survives steps of elimination of iterated dominance. We show that iterated dominance based on mixed strategies is (round-for-round) equivalent to iterated grandfather dominance. 

The equivalence between iterated dominance and iterated grandfather dominance may be useful beyond delivering the main result. In particular, it provides a method for computing the $m$-undominated mixed strategies by searching for improvements relative to all $(m-1)$-undominated pure strategies instead of (what will be) the larger set of all $(m-1)$-undominated mixed strategies. The negative is that grandfather dominance is not a self-referential procedure, in that it references an alternate elimination procedure. Thus, it first requires computing an alternate procedure. 

With the above in mind, we introduce an alternative elimination procedure, termed iterated support dominance. Round-for-round, the procedure is equivalent to iterated dominance and iterated grandfather dominance. Much like iterated grandfather dominance, it only requires searching for improvements relative to a set of the opponents' pure strategies. Unlike grandfather dominance, the procedure is self-referential, in that the relevant set of pure strategies is computed from iterated support dominance itself. 

\section{Setup and definitions}

Let $\Omega$ be a compact metric space. Write $\Delta(\Omega)$ for the set of Borel probability measures on $\Omega$. Endow $\Delta(\Omega)$ with the topology of weak convergence, so that it is a compact metric space. With abuse of notation, if $E\subseteq \Omega$ is Borel, write $\Delta(E)$ for the set $\{ \mu \in \Delta(\Omega): \mu(E)=1 \}$. If $E \subseteq \Omega$ is closed, $\Delta(E)=\{ \mu \in \Delta(\Omega): \mu(E)=1 \}$ is a closed subset of a compact metric set and so compact. Write $\delta_{\omega} \in \Delta(\Omega)$ for the probability measure that assigns probability 1 to $\omega$.  

Given a collection of sets $(\Omega_{i}: i =1,\ldots, I)$, write $\Omega_{-i} = \prod_{j \neq i}\Omega_{j}$ and $\Omega = \prod_{i =1}^{I} \Omega_{i} $. If each $\Omega_{i}$ is a compact metric space, so are $\Omega_{-i}$ and $\Omega$. 

\paragraph{Game} Fix a game $G=(S_{i},u_{i} \colon i=1,\ldots, I)$, where $S_{i}$ is player $i$'s set of pure-strategies and $u_{i}: S_{i} \times S_{-i} \rightarrow \mathbb{R}$ is player $i$'s payoff function. Throughout, we focus on compact and continuous games, i.e., games where each $S_{i}$ is a compact metric space and each $u_{i}$ is continuous. 


Extend $u_{i}$ to the domain $\prod_{j=1}^{I}\Delta(S_{j})$ in the usual way, writing $u_{i}(\sigma_{i},\sigma_{-i})$ for the expected payoff from the mixed strategy $\sigma_{i} \in \Delta(S_{i})$ given the mixed strategy profile $\sigma_{-i} \in \prod_{j \neq i}\Delta(S_{j})$. Similarly, extend $u_{i}$ to the domain $\Delta(S_{i}) \times S_{-i}$.
Note, $u_{i}: \prod_{j =1}^{I}\Delta(S_{j}) \rightarrow \mathbb{R}$ and $u_{i}: \Delta(S_{i}) \times S_{-i}  \rightarrow \mathbb{R}$ are continuous.

\paragraph{Iterated Elimination of Dominated Strategies} We will look at two procedures that iteratively eliminate dominated strategies: one that eliminates pure strategies and one that eliminates mixed strategies. The former involves subsets of pure strategies $S_{i}$ and the latter involves subsets of mixed strategies $\Delta(S_{i})$. To avoid confusion, we will use the following convention: For each $i$, $X_{i}$ will denote a subset of $S_{i}$ and $\Xi_{i}$ will denote a subset of $\Delta(S_{i})$. Note, $X_{-i} = \prod_{j \neq i} X_{j}$ and $\Xi_{-i} = \prod_{j \neq i} \Xi_{j}$ will denote the respective product sets. 

We first define the dominance procedure based on pure strategies.  

\begin{definition}
Fix a non-empty set  $X_{i} \times X_{-i} \subseteq S_{i} \times S_{-i}$. A pure-strategy $s_{i}$ is \textbf{dominated} given $X_{i} \times X_{-i}$ if $s_{i}\in X_{i}$ and there exists some $\sigma_{i} \in \Delta(X_{i})$ so that 
\[
	u_{i}(\sigma_{i},s_{-i}) > u_{i}(s_{i},s_{-i}) \qquad \text{ for all } s_{-i} \in X_{-i}.
\] 
\end{definition}
\noindent Inductively define sets $S_{i}^{m}$ as follows: $S_{i}^{0}=S_{i}$ and $S_{i}^{m+1}$ is the set of $s_{i} \in S_{i}^{m}$ so that $s_{i}$ is not dominated given $S_{i}^{m} \times S_{-i}^{m}$. The set $S^{m} =\prod_{i=1}^{I} S_{i}^{m}$ denotes the set of \textbf{pure} $\bf{m}$\textbf{-undominated} strategy profiles.

We next define the dominance procedure based on mixed strategies.

\begin{definition}
Fix a non-empty set $\Xi_{i} \times \Xi_{-i} \subseteq \Delta(S_{i}) \times \prod_{j \neq i} \Delta(S_{j})$. A mixed-strategy $\sigma_{i}\in \Delta(S_{i})$ is \textbf{dominated} given $\Xi_{i} \times \Xi_{-i} \subseteq \Delta(S_{i}) \times \prod_{j \neq i} \Delta(S_{j})$ if $\sigma_{i}\in \Xi_{i}$ and there exists some $\rho_{i} \in \Xi_{i}$ so that 
\[
	u_{i}(\rho_{i},\sigma_{-i}) > u_{i}(\sigma_{i},\sigma_{-i}) \qquad \text{ for all } \sigma_{-i} \in \Xi_{-i}.
\] 
\end{definition}
\noindent Inductively define sets $\Sigma_{i}^{m}$ as follows: $\Sigma_{i}^{0}=\Delta(S_{i})$ and $\Sigma_{i}^{m+1}$ is the set of $\sigma_{i} \in \Sigma_{i}^{m}$ so that $\sigma_{i}$ is not dominated given $\Sigma_{i}^{m} \times \Sigma_{-i}^{m}$.  The set $\Sigma^{m} =\prod_{i=1}^{I} \Sigma_{i}^{m}$ denotes the set of \textbf{mixed}  $\bf{m}$\textbf{-undominated} strategy profiles.
 %

\paragraph{Termination} Because our focus is on compact and continuous games, $S_{i}^{\infty} := \cap_{m} S_{i}^{m}$ and $\Sigma_{i}^{\infty}:= \cap_{m} \Sigma_i^{m}$ are non-empty. (This is an implication of \citepos{glicksberg1952further} Theorem.) Moreover, each of these dominance procedures stop shrinking at the first limit ordinal: That is, if $s_{i} \in S_{i}^{\infty} $ (resp.\ $\sigma_{i} \in \Sigma_{i}^{\infty} $), then $s_{i}$ (resp.\ $\sigma_{i}$) is undominated given $S_{i}^{\infty} \times S_{-i}^{\infty}$ ($\Sigma_{i}^{\infty} \times \Sigma_{-i}^{\infty}$).\footnote{\citet{arieli2010rationalizability} proves a related result, dealing with rationalizability when players choose pure strategies.} (See Lemma \ref{lem:limit-ordinal}.) With this in mind, the focus will be on stabilization after a finite number of rounds. 

Say that the dominance procedure based on pure (resp.\ mixed) strategies \textbf{terminates} if there is some finite $M$ such that $S^m = S^M$ (resp.\ $\Sigma^m = \Sigma^M$) for all $m \geq M$. If the procedure terminates, then there is a finite $M$ such that either 
\begin{enumerate*}
	\item $M=0$ and $S^{m}=S$ (resp.\ $\Sigma^{m}=\Sigma^0$) for all $m \geq 0$, or 
	\item $S^{M} \subsetneq S^{M-1}$ and $S^{m} = S^{M}$ for all $m \geq M$ (resp.\ $\Sigma^{M} \subsetneq \Sigma^{M-1}$ and $\Sigma^m = \Sigma^M$ for all $m \geq M$). 
\end{enumerate*}
In that case, say that the procedure \textbf{terminates at $M$}.

When the game is finite, the procedure based on pure strategies terminates. The question is whether, in that case, the procedure based on mixed strategies also terminates. More generally, does termination of the procedure based on pure strategies imply termination of the procedure based on mixed strategies?

\section{\label{sec:motivating}Motivating example}

Consider the two-player game, described by the following payoff bi-matrix.  

\begin{table}[h]
    \centering
    \setlength{\extrarowheight}{2pt}
    \begin{tabular}{cc|c|c|c|}
      & \multicolumn{1}{c}{} & \multicolumn{1}{c}{$L$}  & \multicolumn{1}{c}{$C$} & \multicolumn{1}{c}{$R$} \\\cline{3-5}
      \multirow{3}*{}  & $U$ & $4,0$ & $0,4$ & $4,0$ \\\cline{3-5}
      & $M$ & $0,4$ & $4,0$ & $4,0$ \\\cline{3-5}
      & $D$ & $2.1,1$ & $2.1,1$ & $2.1,0$ \\\cline{3-5}
    \end{tabular}
\end{table}
\noindent The first entry in each cell is the payoff to Player 1 and the second entry is the payoff to Player 2. We will show that iterated elimination of mixed strategies does not terminate at $M=1$, even though iterated elimination of pure strategies does. 

Start with iteratively eliminating dominated pure strategies. Observe that
\[
	S_{1}^{1}=S_{1} \qquad \text{ and } \qquad S_{2}^{1}= \{ L,C \}.
\]
In fact, for each $m \geq 1$, $S^{m} =S^{1}$. Thus, this procedure terminates at $M=1$.

Iteratively eliminating dominated mixed strategies is more subtle. On round one, 
\[
	\Sigma_{1}^{1}=\Delta(S_{1}) \qquad \text{ and } \qquad \Sigma_{2}^{1}= \{ \sigma_{2} \in \Delta(S_{2}): \sigma_{2}(R)=0 \}.
\]
To see this, first observe that each $\sigma_{1} \in \Delta(S_{1})$ is undominated given $\Delta(S_{1}) \times \{ \delta_{L}, \delta_{C}, \delta_{R} \}$. In particular, suppose there exists $\rho_{1} \in \Delta(S_{1})$ so that $u_{1}(\rho_{1},L) - u_{1}(\sigma_{1},L)>0$ and $u_{1}(\rho_{1},C) - u_{1}(\sigma_{1},C)>0$. Then 
\[
 	u_{1}(\rho_{1},L) + u_{1}(\rho_{1},C) - u_{1}(\sigma_{1},L) - u_{1}(\sigma_{1},C) = .2(\rho_{1}(D)- \sigma_{1}(D)) >0
\]
and, so, $\rho_{1}(D)>\sigma_{1}(D)$. But then 
\[
	u_{1}(\rho_{1},R) - u_{1}(\sigma_{1},R) = 1.9 \left( \sigma_{1}(D) - \rho_{1}(D)  \right) <0,
\]
and so $\rho_{1}$ does not dominate $\sigma_{1}$. Since $\sigma_{1} \in \Delta(S_{1})$ is undominated given $\Delta(S_{1}) \times \{ \delta_{L}, \delta_{C}, \delta_{R} \}$, it is also undominated given $\Delta(S_{1}) \times \Delta(S_{2})$. An analogous argument establishes that, if $\sigma_{2} \in \Delta(S_{2})$ with $\sigma_{2}(R)=0$, then $\sigma_{2}$ is undominated given $\Delta(S_{2}) \times \{ \delta_{U},\delta_{M},\delta_{D} \} $ and so undominated given $\Delta(S_{2}) \times \Delta(S_{1})$. But, $\sigma_{2}\in \Delta(S_{2})$ with $\sigma_{2}(R)>0$ is dominated given $\Delta(S_{2}) \times \Delta(S_{1})$: Take $\rho_{2} \in \Delta(S_{2})$ so that it redistributes the weight $\sigma_{2}$ assigns to $R$ to both $L$ and $C$, e.g., $\rho_{2}(L)=\sigma_{2}(L)+.5\sigma_{2}(R)$ and $\rho_{2}(C)=\sigma_{2}(C)+.5\sigma_{2}(R)$. Then, for each $\sigma_{1}$, $u_{2}(\rho_{2},\sigma_{1})> u_{2}(\sigma_{2},\sigma_{1})$. 

The iterated elimination procedure based on mixed strategies does not terminate at $M=1$. In fact, if $\sigma_{1} \in \Sigma_{1}^{1}$ has $\sigma_{1}(U),\sigma_{1}(M)>0$, then $\sigma_{1}$ is dominated given $\Sigma_{1}^{1} \times \Sigma_{2}^{1}$; that is, if $\sigma_{1}$ assigns strictly positive probability to both $U$ and $M$, it is dominated given $\Sigma_{1}^{1} \times \Sigma_{2}^{1}$. To see this, consider such a strategy $\sigma_{1}$ and let $\epsilon = \min\{ \sigma_{1}(U),\sigma_{1}(M) \}>0$. Construct $\rho_{1}$ to redistribute weight from both $U$ and $M$ to $D$, i.e., so that 
\[
	\rho_{1}(U) =  \sigma_{1}(U) - \epsilon \qquad \rho_{1}(M) =  \sigma_{1}(M) - \epsilon \qquad \rho_{1}(D) =  \sigma_{1}(D) + 2 \epsilon.
\]
Then, for any $\sigma_{2} \in \Sigma_{2}^{1}=\{ \sigma_{2} \in \Delta(S_{2}): \sigma_{2}(R)=0 \}$, 
\begin{align*}
	u_{1}(\rho_{1},\sigma_{2}) &= \sigma_{2}(L)\left( 4 \rho_{1}(U) + 2.1\rho_{1}(D) \right) + \sigma_{2}(C)\left( 4 \rho_{1}(M) + 2.1\rho_{1}(D) \right) \\
	&= \sigma_{2}(L)\left( 4 \sigma_{1}(U) + 2.1\sigma_{1}(D) + .2 \epsilon \right) + \sigma_{2}(C)\left( 4 \sigma_{1}(M) + 2.1\sigma_{1}(D) + .2 \epsilon \right) \\
	&= u_{1}(\sigma_{1},\sigma_{2}) + .2 \epsilon \\
	&> u_{1}(\sigma_{1},\sigma_{2}),
\end{align*}
as desired.

No other strategy $\sigma_{1}$ is dominated given $\Sigma_{1}^{1} \times \Sigma_{2}^{1}$. To show this, it suffices to show that, if $\sigma_{1}(U) \sigma_{1}(M)=0$, then $\sigma_{1}$ is not dominated given $\Sigma_{1}^{1} \times \{ \delta_{L},\delta_{C} \}$. Focus on the case where $\sigma_{1}(U)=0$, so that $\sigma_{1}(M)=1-\sigma_{1}(D)$. Since $\sigma_{1}(U)=0$,
\[
u_{1}(\rho_{1},L) - u_{1}(\sigma_{1},L)>0 \qquad \text{ if and only if } \qquad 4 \rho_{1}(U) >  2.1(\sigma_{1}(D)- \rho_{1}(D) )
\]
and 
\[
u_{1}(\rho_{1},C) - u_{1}(\sigma_{1},C)>0 \qquad \text{ if and only if } \qquad  1.9( \sigma_{1}(D) - \rho_{1}(D) ) > 4 \rho_{1}(U) .
\]
Since $\rho_{1}(U) \geq 0$, the two requirements are inconsistent. Thus, $\sigma_{1}$ is undominated given $\Sigma_{1}^{1} \times \{ \delta_{L},\delta_{C} \}$.

With the above in mind, round two of the iterated elimination procedure based on mixed strategies gives
\[
	\Sigma_{1}^{2}= \{ \sigma_{1} \in \Sigma_{1}^{1}: \sigma_{1}(U)\sigma_1(M)=0 \}  \qquad \text{ and } \qquad \Sigma_{2}^{2}=\Sigma_{2}^{1}.
\]
Now observe that each $\sigma_{2} \in \Sigma_{2}^{2}$ is undominated given $\Sigma_{2} \times \{ \delta_{U},\delta_{M} \}$ and, so, undominated given $\Sigma_{2} \times \Sigma_{1}$. Thus, for each $m \geq 2$, $\Sigma^{m}=\Sigma^{2}$. So the procedure that eliminates mixed strategies terminates at $M=2$.

Observe that the procedure based on mixed strategies eliminates all the mixed strategies that place strictly positive probability on $U$ and $M$, without eliminating any pure strategies of player 1. There is, as we shall see, a connection between the elimination of pure and mixed strategies, but it does not reduce to eliminating strategies played with positive probability in a dominated mixed strategy. 

%

\section{\label{sec:main}Main result}

In the motivating example, iterated elimination of mixed strategies terminated in two steps. This raises two interrelated questions. First, it leaves open the hypothesis that, when the procedure based on pure strategies terminates, then the procedure based on mixed strategies also terminates. Second, it raises the question of how many additional steps may be required compared to the iterated dominance procedure based on pure strategies. The main theorem addresses both questions. 

%

\begin{theorem}\label{thm:main} 
Let $G$ be a compact and continuous game. If the dominance procedure based on pure strategies terminates at $M$, then the dominance procedure based on mixed strategies terminates at either $M$ or $M+1$. 
\end{theorem}

\begin{remark}
If the procedure based on mixed strategies terminates in finitely many rounds, then so does the procedure based on pure strategies. So Theorem~\ref{thm:main} implies that one procedure terminates if and only if the other terminates.
  \end{remark}

The remainder of the section is devoted to showing Theorem \ref{thm:main}. All subsequent results presume the game is compact and continuous. 

\subsection{\label{subsec:auxiliary}An auxiliary elimination procedure}

To prove Theorem~\ref{thm:main}, it will be convenient to consider an alternate procedure that iteratively eliminates mixed strategies. The new procedure will serve to eliminate mixed strategies based on how they perform relative to pure strategies chosen by other players. First we define the elimination procedure and then we discuss why the procedure suffices.  

\begin{definition}
Fix $X_{i} \times X_{-i} \subseteq S_{i} \times S_{-i}$. Say $\sigma_{i}$ is \textbf{grandfather-dominated} given $X_{i} \times X_{-i}$ if $\sigma_{i} \in \Delta(X_{i})$ and there exists $\rho_{i} \in \Delta(X_{i})$ so that 
\[
	u_{i}(\rho_{i},s_{-i}) > u_{i}(\sigma_{i},s_{-i}) \qquad \text{for each } s_{-i} \in X_{-i}.
\]
\end{definition}
\noindent A mixed strategy $\sigma_{i}$ is grandfather-dominated if it is dominated relative to a ``grandfather'' set of pure strategies---that is, if there is another mixed strategy $\rho_{i}$ that assigns probability 1 to the ``grandfather" set $X_{i}$, so that $i$'s expected payoff from $\rho_{i}$ is higher than from $\sigma_{i}$ given every pure strategy profile $s_{-i}$ in the ``grandfather'' set $X_{-i}$. 

Set $\grand_{i}^{0} = \Delta(S_{i})$ and, for each $m \geq 0$, let $\grand_{i}^{m+1} $ be the set of $\sigma_{i} \in \grand_{i}^{m} $ that are not grandfather dominated given $S_{i}^{m} \times S_{-i}^{m}$. So, $\grand^{m} = \prod_{i=1}^{I}\grand^m_i$ is the set of $\bf{m}$\textbf{-grandfather undominated} strategy profiles.

\begin{proposition}
\label{pro:grandfather}
For each $m$, $\Sigma^{m} = \grand^{m} $.
\end{proposition}
\noindent Proposition \ref{pro:grandfather} says that the $m$-undominated mixed strategies coincide with the $m$-grandfather undominated strategies. 

To understand the result, return to the motivating example and consider a strategy $\sigma_{1} \in \Sigma_{1}^{1}$. We saw that, if $\sigma_{1}(U)=0$, then $\sigma_{1}$ is undominated given $\Delta(S_{1}) \times \{ \delta_{L},\delta_{C} \} $ or, equivalently, not grandfather dominated given $S_{1} \times \{ L,C \}$; this sufficed to conclude that $\sigma_{1}$ is undominated given $\Sigma_{1}^{1} \times \Sigma_{2}^{1}$. The key was that $L$ and $C$ were the only pure-strategies that could get strictly positive probability under (a strategy in) $\Sigma_{2}^{1}$. Similarly, we saw that, if $\sigma_{1}(U)\sigma_{1}(M)>0$, then $\sigma_{1}$ is dominated given $\Sigma_{1}^{1} \times \Sigma_{2}^{1}$. This followed from the fact that there was a $\rho_{1}$ that improved upon $\sigma_{1}$ for both $L$ and $C$, i.e., for the two pure strategies that can get strictly positive probability under (a strategy in) $\Sigma_{2}^{1}$. Thus, $\sigma_{1}$ is grandfather dominated given $S_{1}^{1} \times S_{2}^{1}$. 

The above paragraph suggests that $\sigma_{1} \in \Sigma_{1}^{1}$ is undominated given $\Sigma_{1}^{1} \times \Sigma_{2}^{1}$ if and only if it is not dominated relative to all pure strategies $s_{2}$ that can get strictly positive probability under a strategy in $ \Sigma_{2}^{1}$, i.e., if and only if it is not dominated relative to all pure strategies that ``support'' $\Sigma_{2}^{1}$. Indeed, in Section \ref{sec:discussion} we will see that this is true. The key is that a pure strategy $s_{2}$ can get strictly positive probability under some $\sigma_{2} \in \Sigma_{2}^{1}$ if and only if it is $1$-undominated for player $2$. 

\subsection{Termination}

Theorem \ref{thm:main} follows immediately from Proposition \ref{pro:grandfather} and the following Lemma: 

\begin{lemma}
\label{lem:grandfather-degenerate}	
A strategy $s_{i} \in S_{i}^{m}$ if and only if $\delta_{s_{i}} \in \Sigma_{i}^{m}$.
\end{lemma}

\begin{proof}[Proof of Theorem \ref{thm:main}]
  Suppose the dominance procedure based on pure strategies terminates at $M$. Then, for each $m \geq M$, $S^{m}=S^{M}$. It follows that, for each $m \geq M+1$, $\grand^{m} =\grand^{M+1}$. Moreover, if $M \geq 1$, then there exists $S_{i}^{M-1} \backslash S_{i}^{M} \neq \emptyset$; in that case, Proposition~\ref{pro:grandfather} and  Lemma~\ref{lem:grandfather-degenerate} imply that $\Sigma_{i}^{M-1} \backslash \Sigma_{i}^{M}=\grand_{i}^{M-1} \backslash \grand_{i}^{M} \neq \emptyset$. 
\end{proof}

\subsection{Proof of the equivalence}

We now turn to the proof of the equivalence between iterated elimination of dominated mixed strategies and iterated elimination of grandfather-dominated mixed strategies. It will be convenient to simultaneously prove Proposition \ref{pro:grandfather}, Lemma \ref{lem:grandfather-degenerate}, and a third hypothesis. This is summarized by the following revised version of Proposition \ref{pro:grandfather}: 

\addtocounter{proposition}{-1}
\renewcommand{\theproposition}{\arabic{proposition}\ensuremath{^*}}
\begin{proposition}
\label{pro:equiv}	
For each $m$,  
\begin{enumerate}
	\item\label{pro:equiv-n1} $\grand_{i}^{m} = \Sigma_{i}^{m}$; 
	\item\label{pro:equiv-n2} $s_{i} \in S_{i}^{m}$ if and only if $\delta_{s_{i}} \in \Sigma_{i}^{m}$; and 
	\item\label{pro:equiv-n3} if $\sigma_{i} \in \Sigma_{i}^{m}$ then $\sigma_{i}(S_{i}^{m})=1$.
\end{enumerate}
\end{proposition}
\renewcommand{\theproposition}{\arabic{proposition}}

We now provide the key steps of the proof. Certain technical steps can be found in the Appendix. 

\begin{proof}[Proof of Proposition \ref{pro:equiv}]
The claim trivially holds for $m=0$. Suppose the three hypotheses hold for $m$; we will show that they hold for $m+1$.

\ \ 

\noindent \emph{Part \ref{pro:equiv-n1}.} By part \ref{pro:equiv-n1} of the induction hypothesis, $\hat{\Sigma}_{i}^{m} = \Sigma_{i}^{m}$. Thus, it suffices to show that
\begin{enumerate*}
	\item if $\sigma_{i} \in \Sigma_{i}^{m} \backslash \Sigma_{i}^{m+1}$, then $\sigma_{i} \not\in \hat{\Sigma}_{i}^{m+1}$, and
	\item if $\sigma_{i} \in \hat{\Sigma}_{i}^{m} \backslash \hat{\Sigma}_{i}^{m+1}$, then $\sigma_{i} \not\in \Sigma_{i}^{m+1}$.
\end{enumerate*}
If so, $\hat{\Sigma}_{i}^{m+1} = \Sigma_{i}^{m+1}$.

First, suppose $\sigma_{i} \in \Sigma_{i}^{m} \backslash \Sigma_{i}^{m+1}$. Then there exists some $\rho_{i} \in \Sigma_{i}^{m}$ so that 
\[
	u_{i}(\rho_{i},\sigma_{-i}) > u_{i}(\sigma_{i},\sigma_{-i}) \qquad \text{ for all } \sigma_{-i} \in \Sigma_{-i}^{m}. 
\]
By part \ref{pro:equiv-n2} of the induction hypothesis, for each $s_{-i} \in S_{-i}^{m}$, $\delta_{s_{-i}} \in \Sigma_{-i}^{m}$. As such, 
\[
	u_{i}(\rho_{i},s_{-i}) > u_{i}(\sigma_{i},s_{-i}) \qquad \text{ for all } s_{-i} \in S_{-i}^{m}. 
\]
By part \ref{pro:equiv-n3} of the induction hypothesis, $\rho_{i}(S_{i}^{m})=1$. Thus, $\sigma_{i}$ is grandfather-dominated given $S_{i}^{m} \times S_{-i}^{m}$, i.e., $\sigma_{i} \in \hat{\Sigma}_{i}^{m} \backslash \hat{\Sigma}_{i}^{m+1}$. 

Next, suppose $\sigma_{i} \in \hat{\Sigma}_{i}^{m} \backslash \hat{\Sigma}_{i}^{m+1}$. Then there exists some $\rho_{i} \in \Delta(S_{i}^{m})$ so that 
\[
	u_{i}(\rho_{i},s_{-i}) > u_{i}(\sigma_{i},s_{-i}) \qquad \text{ for all } s_{-i} \in S_{-i}^{m}. 
\]
Fix some $\sigma_{-i} = (\sigma_{j}: j \neq i) \in \Sigma_{-i}^{m}$ and note that, for each $j$, $\sigma_{j}(S_{j}^{m})=1$. (This uses part \ref{pro:equiv-n3} of the induction hypothesis.) Thus, for each $\sigma_{-i} \in \Sigma_{-i}^{m}$ and each associated product measure $\hat{\sigma}_{-i} = \otimes_{j \neq i} \sigma_{j}$, 
\[
	u_{i}(\rho_{i},\sigma_{-i}) = \int_{S_{-i}^{m}} u_{i}(\rho_{i},s_{-i})d \hat{\sigma}_{-i} >  \int_{S_{-i}^{m}} u_{i}(\sigma_{i},s_{-i})d \hat{\sigma}_{-i} = u_{i}(\sigma_{i},\sigma_{-i}) \qquad \text{ for all } \sigma_{-i} \in \Sigma_{-i}^{m}. 
\]
Thus, $\sigma_{i}$ is dominated given $\Sigma_{i}^{0} \times \Sigma_{-i}^{m}$. By Lemma \ref{lem:all-dominated}, $\sigma_{i}$ is dominated given $\Sigma_{i}^{m} \times \Sigma_{-i}^{m}$.

\ \ 

\noindent \emph{Part \ref{pro:equiv-n2}.} Fix $s_{i} \in S_{i}^{m}$. By part \ref{pro:equiv-n2} of the induction hypothesis, $\delta_{s_{i}} \in \Sigma_{i}^{m}$. So, it suffices to show that $\delta_{s_{i}} \in \Sigma_{i}^{m} \backslash \Sigma_{i}^{m+1}$ if and only if $s_{i} \in S_{i}^{m} \backslash S_{i}^{m+1}$. 

First suppose $\delta_{s_{i}} \in \Sigma_{i}^{m} \backslash \Sigma_{i}^{m+1}$. Then there exists a $\sigma_{i} \in \Sigma_{i}^{m}$ so that 
\[
	u_{i}(\sigma_{i},\sigma_{-i}) > u_{i}(\delta_{s_{i}},\sigma_{-i}) \qquad \text{ for all } \sigma_{-i} \in \Sigma_{-i}^{m}. 
\]
Again applying part \ref{pro:equiv-n2} of the induction hypothesis, for each $s_{-i} \in S_{-i}^{m}$, $\delta_{s_{-i}} \in \Sigma_{-i}^{m}$ and, so,  
\[
	u_{i}(\sigma_{i},s_{-i}) > u_{i}(\delta_{s_{i}},s_{-i}) = u_{i}(s_{i},s_{-i}) \qquad \text{ for all } s_{-i} \in S_{-i}^{m}. 
\]
Finally observe that part \ref{pro:equiv-n3} of the induction hypothesis gives $\sigma_{i}(S_{i}^{m})=1$. Thus, $s_{i}$ is dominated given $S_{i}^{m} \times S_{-i}^{m}$, i.e, $s_{i} \in S_{i}^{m} \backslash S_{i}^{m+1}$. 

%

Next suppose $s_{i} \in S_{i}^{m} \backslash S_{i}^{m+1}$. Then there exists a $\sigma_{i} \in \Delta(S_{i}^{m})$ so that 
\[
	u_{i}(\sigma_{i},s_{-i}) > u_{i}(s_{i},s_{-i})  \qquad \text{ for all } s_{-i} \in S_{-i}^{m}. 
\]
Fix $\sigma_{-i} = (\sigma_{j}: j \neq i) \in \Sigma_{-i}^{m}$ and note that, for each $j$, $\sigma_{j}(S_{j}^{m})=1$. (This uses part \ref{pro:equiv-n3} of the induction hypothesis.) Thus, for each $\sigma_{-i} \in \Sigma_{-i}^{m}$ and each associated product measure $\hat{\sigma}_{-i} = \otimes_{j \neq i} \sigma_{j}$, 
\[
	u_{i}(\sigma_{i},\sigma_{-i}) = \int_{S_{-i}^{m}} u_{i}(\sigma_{i},s_{-i})d \hat{\sigma}_{-i} >  \int_{S_{-i}^{m}} u_{i}(\delta_{s_{i}},s_{-i})d \hat{\sigma}_{-i} = u_{i}(\delta_{s_{i}},\sigma_{-i}) \qquad \text{ for all } \sigma_{-i} \in \Sigma_{-i}^{m}. 
\]
Thus, $\delta_{s_{i}}$ is dominated given $\Sigma_{i}^{0} \times \Sigma_{-i}^{m}$. By Lemma \ref{lem:all-dominated}, $\delta_{s_{i}}$ is dominated given $\Sigma_{i}^{m} \times \Sigma_{-i}^{m}$, i.e., $\delta_{s_{i}} \not\in \Sigma_{i}^{m+1}$.

\ \ 

\noindent \emph{Part \ref{pro:equiv-n3}.} Fix $\sigma_{i} \in \Sigma_{i}^{m+1}$. By part \ref{pro:equiv-n3} of the induction hypothesis, $\sigma_{i}(S_{i}^{m})=1$. Suppose, contra hypothesis, that $\sigma_{i}$ does not assign probability 1 to $S_{i}^{m+1}$. By Lemma \ref{lem:m-closed}, $S_{i}^{m}\backslash S_{i}^{m+1}$ is Borel and so $\sigma_{i}(S_{i}^{m}\backslash S_{i}^{m+1})>0$. We will construct some $\nu_{i} \in \Delta(S_{i}^{m})$ so that 
\[
	u_{i}(\nu_{i},s_{-i}) > u_{i}(\sigma_{i},s_{-i}) \qquad \text{ for all } s_{-i} \in S_{-i}^{m}.
\]
Thus, $\sigma_{i}$ is grandfather dominated given $S_{i}^{m} \times S_{-i}^{m}$ and, so, by part \ref{pro:equiv-n1} for $m+1$ (which we just proved), $\sigma_{i} \not\in \Sigma_{i}^{m+1}$. 

By Lemma \ref{lem:measurable-function}, there exists a measurable function $f_{i}: S_{i}^{m} \rightarrow \Delta(S_{i}^{m})$ (a ``Kernel'') with the properties that
\begin{enumerate}
	\item for each $s_{i} \in S_{i}^{m+1}$, $f_{i}(s_{i}) = \delta_{s_{i}}$, and 
	\item\label{number-fi} for each $s_{i} \in S_{i}^{m} \backslash S_{i}^{m+1}$, $f_{i}(s_{i})$ satisfies $u_{i} (f_{i} (s_{i}), s_{-i}) > u_{i} (s_{i}, s_{-i})$ for all $s_{-i} \in S_{-i}^{m}$.
\end{enumerate}

For each Borel $E_{i} \subseteq S_{i}^{m}$, there is a mapping $g[E_{i}]: S_{i}^{m} \rightarrow [0,1]$ so that $g[E_{i}](s_{i}) = f_{i}(s_{i})(E_{i})$. Note, each $g[E_{i}]$ is Borel. To see this, note that for any Borel $B \subseteq [0,1]$, $\{ \sigma_{i} \in \Delta(S_{i}^{m}): \sigma_{i}(E_{i}) \in B \}$ is Borel \citep[Theorem 15.13]{aliprantis2007ida}; since $f_{i}$ is measurable,  
\[
	g[E_{i}]^{-1}(B) =  f_{i}^{-1}(\{ \sigma_{i} \in \Delta(S_{i}^{m}): \sigma_{i}(E_{i}) \in B \})
\]
is Borel. 

Now define $\nu_{i}$ so that, for each Borel $E_{i} \subseteq S_{i}^{m}$, 
\[
	\nu_{i}(E_{i}) = \int_{S_{i}^{m}} g[E_{i}](s_{i}) d\sigma_{i}(s_{i}).
\]
This forms a probability measure. In particular, $g[\emptyset]$ (resp.\ $g[S_{i}^{m}]$) is constant on $0$ (resp.\ $1$) and so $\nu_{i}(\emptyset)=0$ (resp.\ $\nu_{i}(S_{i}^{m})=1$). Moreover, given pairwise disjoint Borel sets $E_{i}^{1},E_{i}^{2},\ldots$, 
\begin{align*}
	\nu_{i}( \bigcup_{k \geq 1} E_{i}^{k}) &= \int_{S_{i}^{m}} f_{i}(s_i)( \bigcup_{k \geq 1} E_{i}^{k})  d \sigma_{i}(s_{i}) \\
	&= \sum_{k \geq 1} \int_{S_{i}^{m}} f_{i}(s_i)(  E_{i}^{k})  d \sigma_{i}(s_{i}) \\
	&= \sum_{k \geq 1} \nu_{i}( E_{i}^{k}),
\end{align*}
and so $\nu_{i}$ is countably additive. 

Now fix some $s_{-i} \in S_{-i}^{m}$ and observe that
\begin{align*}
	\int_{S_{i}^{m}} u_{i}(s_{i},s_{-i}) d\nu_{i}(s_i) &= \int_{S_{i}^{m+1} } u_{i}(s_{i},s_{-i}) d\nu_{i}(s_i)  + \int_{S_{i}^{m}\backslash S_{i}^{m+1}} u_{i}(s_{i},s_{-i}) d\nu_{i}(s_i)  \\
	&= \int_{S_{i}^{m+1} } \int_{S_{i}^{m}} u_{i}(r_{i},s_{-i}) d\delta_{s_{i}}(r_i) d\sigma_{i}(s_i)  + \int_{S_{i}^{m}\backslash S_{i}^{m+1}} \int_{S_{i}^{m}} u_{i}(r_{i},s_{-i}) d(f_{i}(s_{i}))(r_i) d\sigma_{i}(s_i) \\
	&= \int_{S_{i}^{m+1} }u_{i}(s_{i},s_{-i}) d\sigma_{i}(s_i)  + \int_{S_{i}^{m}\backslash S_{i}^{m+1}} \int_{S_{i}^{m}} u_{i}(r_{i},s_{-i}) d(f_{i}(s_{i}))(r_i) d\sigma_{i}(s_i) \\
	&> \int_{S_{i}^{m+1} }u_{i}(s_{i},s_{-i}) d\sigma_{i}(s_i)  + \int_{S_{i}^{m}\backslash S_{i}^{m+1}} u_{i}(s_{i},s_{-i}) d\sigma_{i}(s_i) \\	
	&= \int_{S_{i}^{m}} u_{i}(s_{i},s_{-i}) d\sigma_{i}(s_i), 
\end{align*}
where the inequality follows from property \ref{number-fi} in how $f_{i}$ was defined.
Thus, $\sigma_{i}$ is grandfather-dominated given $S_{i}^{m} \times S_{-i}^{m}$ and, so, by part \ref{pro:equiv-n1}, $\sigma_{i} \not\in \Sigma_{i}^{m+1}$, as desired.
\end{proof}

\section{\label{sec:discussion}Support Dominance}

Proposition \ref{pro:grandfather} is arguably of interest beyond its role in proving Theorem \ref{thm:main}. In particular, it yields a method for computing $m$-undominated mixed strategies that requires searching for improvements only relative to the $(m-1)$-undominated pure strategies, $S_{-i}^{m-1}$, rather than the larger set of $(m-1)$-undominated mixed strategies, $\Sigma_{-i}^{m-1}$. 

We now present an alternate characterization of $m$-undominated mixed strategies, based on a concept that we will call support dominance. Like the procedure based on grandfather dominance, it requires searching for improvements relative to pure strategies of the other players. Unlike the procedure based on grandfather dominance, it is self-referential, not requiring the analyst to first compute an alternate procedure. Thus, in certain applications, it may be of use. 

Fix some $ \Xi_{-i} = \prod_{j \neq i} \Xi_{j}  \subseteq \prod_{j \neq i} \Delta(S_{j})$. Say $s_{-i}=(s_{j}: j \neq i)$ \textbf{supports} $\Xi_{-i}$ if there exists some $\sigma_{-i} = (\sigma_{j}: j \neq i) \in \Xi_{-i}$ with $\sigma_{j}(s_{j})>0$ for each $j \neq i$. Write $\supp(\Xi_{-i})$ for the set of strategies that support $\Xi_{-i}$. In the example, $\supp(\Sigma_{2}^{1}) = \{ L,C \}$.

\begin{definition}
Say $\sigma_{i}$ is \textbf{support-dominated} given $\Xi_{i} \times \Xi_{-i} \subseteq \Delta(S_{i}) \times \prod_{j \neq i} \Delta(S_{j})$ if $\sigma_{i}\in \Xi_{i}$ and there exists some $\rho_{i} \in \Xi_{i}$ so that 
\[
	u_{i}(\rho_{i},s_{-i}) > u_{i}(\sigma_{i},s_{-i}) \qquad \text{ for all } s_{-i} \in \supp(\Xi_{-i}).
\] 
\end{definition}
\noindent Inductively define sets $\spdom_{i}^{m}$ and $\spdom_{-i}^{m}$ as follows: $\spdom_{i}^{0}=\Delta(S_{i})$ and $\spdom_{-i}^{0}=\prod_{j \neq i} \Delta(S_{j})$. Then, $\spdom_{i}^{m+1}$ is the set of $\sigma_{i} \in \spdom_{i}^{m}$ so that $\sigma_{i}$ is not support-dominated given $\spdom_{i}^{m} \times \spdom_{-i}^{m}$ and $\spdom_{-i}^{m+1} = \prod_{j \neq i} \spdom_{j}^{m+1}$. The set $\spdom^{m} =\prod_{i=1}^{I} \spdom_{i}^{m}$ denotes the set of  $\bf{m}$\textbf{-support-undominated} strategy profiles.

\begin{proposition}
\label{pro:mixed-weird}
For each $m$, $\Sigma^{m} = \grand^{m} = \spdom^{m}$.
\end{proposition}

Section \ref{subsec:auxiliary} provided intuition for this result, based on the motivating example. In particular, there we argued that $\sigma_{1} \in \Sigma_{1}^{1}$ is undominated given $\Sigma_{1}^{1} \times \Sigma_{2}^{1}$ if and only if it is not dominated relative to all pure strategies $s_{2}$ that can get strictly positive probability under a strategy in $ \Sigma_{2}^{1}$---i.e., if and only if it is not support dominated given $\Sigma_{1}^{1} \times \Sigma_{2}^{1}$. 

\begin{proof}[Proof of Proposition \ref{pro:mixed-weird}]
The proof is by induction on $m$. The case of $m=0$ is immediate. So suppose the claim holds for $m$. We will show that $\spdom_{i}^{m+1} = \grand_{i}^{m+1}  $. Then, by Proposition \ref{pro:grandfather}, the claim follows. 

First, fix $\sigma_{i} \in \spdom_{i}^{m} = \grand_{i}^{m} $ and suppose $\sigma_{i} \not\in \spdom_{i}^{m+1} $. Then there exists some $\rho_{i} \in \spdom_{i}^{m} $ so that
\[
	u_{i}(\rho_{i},s_{-i}) > u_{i}(\sigma_{i},s_{-i}) \qquad \text{ for all } s_{-i} \in \supp(\spdom_{-i}^{m}).
\]
By the induction hypothesis,  $\rho_{i} \in \Sigma_{i}^{m}$ and so, by Proposition \ref{pro:equiv}\ref{pro:equiv-n3}, $\rho_{i}(S_{i}^{m})=1$. Similarly, by the induction hypothesis, $\supp(\spdom_{-i}^{m}) = \supp(\Sigma_{-i}^{m})$ and, so, by Proposition \ref{pro:equiv}\ref{pro:equiv-n2}-\ref{pro:equiv-n3}, $\supp(\Sigma_{-i}^{m}) = S_{-i}^{m}$. Thus, $\sigma_{i}$ is grandfather-dominated given $S_{i}^{m} \times S_{-i}^{m}$ and so $\sigma_{i} \not\in \grand_{i}^{m+1}$. 

For the converse, fix $\sigma_{i} \in \grand_{i}^{m} =\spdom_{i}^{m} $ and suppose $\sigma_{i} \not\in \grand_{i}^{m+1} $. By the induction hypothesis, $\sigma_{i} \in \Sigma_{i}^{m} \backslash \Sigma_{i}^{m+1}$. So there exists some $\rho_{i} \in \Sigma_{i}^{m} $ so that
\[
	u_{i}(\rho_{i},\sigma_{-i}) > u_{i}(\sigma_{i},\sigma_{-i}) \qquad \text{ for all } \sigma_{-i} \in \Sigma_{-i}^{m}.
\]
By Proposition \ref{pro:equiv}\ref{pro:equiv-n2}, 
\[
	u_{i}(\rho_{i},s_{-i}) > u_{i}(\sigma_{i},s_{-i}) \qquad \text{ for all } s_{-i} \in S_{-i}^{m}.
\]
By the induction hypothesis, $\rho_{i} \in \spdom_{i}^{m} $. Again applying the induction hypothesis, $\supp(\spdom_{-i}^{m}) = \supp(\Sigma_{-i}^{m})$; so, by Proposition \ref{pro:equiv}\ref{pro:equiv-n2}-\ref{pro:equiv-n3}, $\supp(\spdom_{-i}^{m}) = \supp(\Sigma_{-i}^{m}) = S_{-i}^{m}$. Thus, $\sigma_{i}$ is support-dominated given $\spdom_{i}^{m} \times \spdom_{-i}^{m}$ and so $\sigma_{i} \not\in \spdom_{i}^{m+1}$. 
\end{proof}

%

\bibliographystyle{apsr}
\bibliography{GT}

\newpage
\setcounter{page}{1}
\appendix

\section{Appendix}

It will be convenient to define $U_{i}^{m} : S_{i}^{m} \times \Delta(S_{i}^{m}) \times S_{-i}^{m} \rightarrow \mathbb{R}$ by
\[
	U_{i}^{m}(s_{i},\sigma_{i},s_{-i}) = u_{i}(\sigma_{i}, s_{-i}) - u_{i}(s_{i}, s_{-i}). 
\]
Note, $U_{i}^{m}$ is continuous. (See, e.g., Theorem 15.5 in \citealp{aliprantis2007ida}.)

\begin{lemma}
\label{lem:m-closed}\mbox{ }
\begin{enumerate}
	\item\label{lem:Sm-closed} For each $i=1,\ldots,I$ and each $m \geq 1$, $S_{i}^{m}$ is closed. 
	\item\label{lem:Sigmam-closed} For each $i=1,\ldots,I$ and each $m \geq 1$, $\Sigma_{i}^{m}$ is closed. 	
\end{enumerate}
\end{lemma}

\begin{proof}
We show part \ref{lem:Sm-closed}. The proof of part \ref{lem:Sigmam-closed}	is effectively a relabeling. 
	
We proceed by induction. By definition $S^0_i$ is closed.  Suppose that, for each $i$, $S_{i}^{m}$ is closed. To show that $S_{i}^{m+1}$ is closed, it suffices to show that $S_{i}^{m} \backslash S_{i}^{m+1}$ is (relatively) open. If so 
\[
	S_{i}^{m+1} =  S_{i}^{m} \cap (S_{i} \backslash (S_{i}^{m} \backslash S_{i}^{m+1}))
\]
is the intersection of two closed sets (since $S_i^m$ is closed, the complement of the relatively open $(S_{i}^{m} \backslash S_{i}^{m+1})$ is closed) and so closed. 


Observe that $\{ (s_{i},\sigma_{i}) \} \times S_{-i}^{m}$ is compact and $U_{i}^{m}$ is continuous. So, each $U_{i}^{m}(s_{i},\sigma_{i},\cdot) : S_{-i}^{m} \rightarrow \mathbb{R}$ achieves a minimum. Thus, there is a value function $v_{i}^{m} : S_{i}^{m} \times \Delta(S_{i}^{m}) \rightarrow \mathbb{R}$ defined by 
\[
	v_{i}^{m}(s_{i},\sigma_{i}) = \min_{s_{-i} \in S_{-i}^{m}}U_{i}^{m}(s_{i},\sigma_{i},s_{-i}) 
\]
and, by Berge's Maximum Theorem, $v_{i}^{m}$ is continuous. Similarly, since each $\{ s_{i} \} \times \Delta(S_{i}^{m})$ is compact and $v_{i}^{m}(s_{i},\cdot): \Delta(S_{i}^{m}) \rightarrow \mathbb{R}$ is continuous, we can define a function $V_{i}^{m} : S_{i}^{m} \rightarrow \mathbb{R}$ by
\[
	V_{i}^{m}(s_{i}) = \max_{\sigma_{i} \in \Delta(S_{i}^{m})}v_{i}^{m}(s_{i},\sigma_{i})
\]
and, by Berge's Maximum Theorem, $V_{i}^{m}$ is continuous. 

To show that $S_{i}^{m} \backslash S_{i}^{m+1}$ is relatively open, it suffices to show that $(V_{i}^{m})^{-1}((0,\infty)) = S_{i}^{m} \backslash S_{i}^{m+1}$: Observe $s_{i} \in S_{i}^{m} \backslash S_{i}^{m+1}$ if and only if $s_{i} \in S_{i}^{m}$ and there exists some $\sigma_{i} \in  \Delta(S_{i}^{m})$ so that $U_{i}^{m}(s_{i},\sigma_{i},s_{-i}) >0$ for each $s_{-i} \in S_{-i}^{m}$. Thus, $s_{i} \in S_{i}^{m} \backslash S_{i}^{m+1}$ if and only if $s_{i} \in S_{i}^{m}$ and there exists some $\sigma_{i} \in  \Delta(S_{i}^{m})$ so that $v_{i}^{m}(s_{i},\sigma_{i}) >0$ or, equivalently, if and only if $V_{i}^{m}(s_{i})>0$.
\end{proof}

\begin{corollary}
For each $i=1,\ldots,I$ and each $m \geq 1$, $S_{i}^{m}$ and $\Sigma_{i}^{m}$ are compact. 	
\end{corollary}

\begin{lemma}
\label{lem:limit-ordinal}	
Let $G$ be a compact and continuous game. 
\begin{enumerate}
	\item\label{lem:limit-ordinal-n1} If $s_{i} \in S_{i}^{\infty} $, then $s_{i} $ is undominated given $S_{i}^{\infty} \times S_{-i}^{\infty} $. 
	\item\label{lem:limit-ordinal-n2} If $\sigma_{i} \in \Sigma_{i}^{\infty} $, then $\sigma_{i} $ is undominated given $\Sigma_{i}^{\infty} \times \Sigma_{-i}^{\infty}$. 
\end{enumerate}
\end{lemma}

\begin{proof}	
We prove part \ref{lem:limit-ordinal-n2}; the proof of part \ref{lem:limit-ordinal-n1} is analogous, up to relabeling. 

Fix $\sigma_{i},\rho_{i} \in \Sigma_{i}^{\infty} $, i.e., for each $m$, $\sigma_{i} \in \Sigma_{i}^{m}$. Then, for each $m$, there exists some $\sigma_{-i}^{m} \in \Sigma_{-i}^{m}$ so that $u_{i}(\sigma_{i},\sigma_{-i}^{m}) - u_{i}(\rho_{i},\sigma_{-i}^{m}) \geq 0$. Consider the sequence $(\sigma_{-i}^{m} : m \geq 0)$ and note that, since $\Sigma_{-i}^{0}$ is compact, there exists a subsequence of $(\sigma_{-i}^{m} : m \geq 0)$ that converges to some $\sigma_{-i}^{*} \in \Sigma_{-i}^{0}$. It suffices to show that, for each $m$, $\sigma_{-i}^{*} \in \Sigma_{-i}^{m}$. If so, by continuity of $u_{i}$, $u_{i}(\sigma_{i},\sigma_{-i}^{*}) - u_{i}(\rho_{i},\sigma_{-i}^{*}) \geq 0$. Repeating the argument for each $\rho_{i} \in \Sigma_{i}^{\infty}$, $\sigma_{i}$ is undominated given $\Sigma_{i}^{\infty} \times \Sigma_{-i}^{\infty} $. 

Fix $m$ and note that, for each $n \geq m$, $\sigma_{-i}^{n} \in \Sigma_{-i}^{n} \subseteq \Sigma_{-i}^{m}$. Since $\Sigma_{-i}^{m}$ is closed (Lemma \ref{lem:m-closed}\ref{lem:Sigmam-closed}), the convergent subsequence of $(\sigma_{-i}^{m} : m \geq 0)$ must converge to a member of $\Sigma_{-i}^{m}$, i.e., $\sigma_{-i}^{*} \in \Sigma_{-i}^{m}$.
\end{proof}

\begin{lemma}
\label{lem:measurable-function}	
There exists a measurable function $f_{i}: S_{i}^{m} \rightarrow \Delta(S_{i}^{m})$ that satisfies the following: 
\begin{enumerate}
	\item If $s_{i} \in S_{i}^{m} \backslash S_{i}^{m+1}$, then $U_{i}^{m}(s_{i},f_{i}(s_{i}), s_{-i}) >0$.
	\item If $s_{i} \in S_{i}^{m+1}$, then $f_{i}(s_{i}) = \delta_{s_{i}}$.
\end{enumerate}
\end{lemma}

\begin{proof}
As in the proof of Lemma \ref{lem:m-closed}, begin by defining a value function $v_{i}^{m} : S_{i}^{m} \times \Delta(S_{i}^{m}) \rightarrow \mathbb{R}$ as
\[
	v_{i}^{m}(s_{i},\sigma_{i}) = \min_{s_{-i} \in S_{-i}^{m}}U_{i}^{m}(s_{i},\sigma_{i},s_{-i}). 
\]
By Berge's Maximum Theorem, $v_{i}^{m}$ is continuous. Define a correspondence $\Phi_{i}^{m} : S_{i}^{m}  \twoheadrightarrow \Delta(S_{i}^{m})$ so that 
\[
	\Phi_{i}^{m}(s_{i}) = \{ \sigma_{i} \in \Delta(S_{i}^{m}) : \sigma_{i} \in \arg \max_{\Delta(S_{i}^{m})} v_{i}^{m}(s_{i},\cdot)  \}. 
\]
Since $\Delta(S_{i}^{m})$ is compact and $v_{i}^{m}$ is continuous, it follows from Berge's Maximum Theorem that $\Phi_{i}^{m}$ is upper hemi-continuous with non-empty closed values. 

In fact, $\Phi_{i}^{m}$ is measurable, i.e., for each closed set $C_{i}^{m} \subseteq \Delta(S_{i}^{m})$, the lower inverse $\{ s_{i} \in S_{i}^{m}: \Phi_{i}^{m}(s_{i}) \cap C_{i}^{m} \neq \emptyset \}$ is Borel. To see this, fix a closed set $C_{i}^{m}$. By upper hemicontinuity, $\{ s_{i} \in S_{i}^{m}: \Phi_{i}^{m}(s_{i}) \subseteq (\Delta(S_{i}^{m}) \backslash C_{i}^{m}) \}$ is open and so 
\[
 	\{ s_{i} \in S_{i}^{m}: \Phi_{i}^{m}(s_{i}) \cap C_{i}^{m} \neq \emptyset \} = S_{i}^{m} \backslash \{ s_{i} \in S_{i}^{m}: \Phi_{i}^{m}(s_{i}) \subseteq (\Delta(S_{i}^{m}) \backslash C_{i}^{m}) \}  
\]
is closed and therefore Borel. 

Since $\Phi_{i}^{m}$ is measurable, it is weakly measurable. (See Lemma 17.2 in \citealp{aliprantis2007ida}.) Thus, by the Kuratowski-Ryall-Nardzewski Selection Theorem, there exists a measurable selection of $\Phi_{i}^{m}$, i.e., a measurable $g_{i} : S_{i}^{m} \rightarrow \Delta(S_{i}^{m}) $ so that, for each $s_{i} \in S^m_{i}$, $g_{i}(s_{i}) \in \Phi_{i}^{m}(s_{i})$. Moreover, for each $s_{i} \in S_{i}^{m} \backslash S_{i}^{m+1}$ and each $s_{-i} \in S_{-i}^{m}$, $U_{i}^{m}(s_{i},g_{i}(s_{i}),s_{-i})>0$. To see this, fix $s_{i} \in S_{i}^{m} \backslash S_{i}^{m+1}$, and note that there exists some $\sigma_{i} \in \Delta(S_{i}^{m})$ with $v_{i}^{m}(s_{i},\sigma_{i}) >0$. Thus, $v_{i}^{m}(s_{i},g_{i}(s_{i}))  = \max_{\Delta(S_{i}^{m})} v_{i}^{m}(s_{i},\cdot) >0$. 

Construct $f_{i} : S_{i}^{m} \rightarrow \Delta(S_{i}^{m})$ so that 
\begin{enumerate*}
	\item if $s_{i} \in S_{i}^{m} \backslash S_{i}^{m+1}$, $f_{i}(s_{i}) = g_{i}(s_{i})$, and 
	\item if $s_{i} \in S_{i}^{m+1}$, then $f_{i}(s_{i}) = \delta_{s_{i}}$.
\end{enumerate*}
Since the mapping $s_{i} \mapsto \delta_{s_{i}}$ is continuous and $S_{i}^{m} \backslash S_{i}^{m+1}$ is (relatively) open (Lemma \ref{lem:m-closed}), $f_{i}$ is measurable.  
\end{proof}

\begin{lemma}
\label{lem:all-dominated}
Let $\sigma_{i} \in \Sigma_{i}^{m}$. If $\sigma_{i}$ is dominated given $\Sigma_{i}^{0} \times \Sigma_{-i}^{m}$, then it is dominated given $\Sigma_{i}^{m} \times \Sigma_{-i}^{m}$.
\end{lemma}

To prove Lemma \ref{lem:all-dominated}, it will be useful to introduce a map $\Upsilon_{i}^{m}:\Delta(S_{i}) \times \Delta(S_{i}) \times \Sigma_{-i}^{m} \rightarrow \mathbb{R}$ so that 
\[
	\Upsilon_{i}^{m}(\sigma_{i},\rho_{i},\sigma_{-i}) = u_{i}(\rho_{i}, \sigma_{-i}) - u_{i}(\sigma_{i}, \sigma_{-i}).
\]
Note, $\Upsilon_{i}^{m}$ is continuous. (See, e.g., Theorem 15.5 in \citealp{aliprantis2007ida}.)

\begin{proof}[Proof of Lemma \ref{lem:all-dominated}]
Define 	
\begin{align*}
\dom_{i} & := \{  \rho_{i} \in \Delta(S_{i}): u_{i}(\rho_{i},\sigma_{-i}) > u_{i}(\sigma_{i},\sigma_{-i}) \text{ for all } \sigma_{-i} \in \Sigma_{-i}^{m} \} \\ 
& = \{  \rho_{i} \in \Delta(S_{i}): \Upsilon_{i}^{m}(\sigma_{i},\rho_{i},\sigma_{-i})>0 \text{ for all } \sigma_{-i} \in \Sigma_{-i}^{m} \}	
\end{align*}
and, for each $k =0,\ldots,m$, $\dom_{i}^{k} := \dom_{i} \cap \Sigma_{i}^{k}$. So, $\dom_{i}^{k}$ is the set of mixed strategies in $\Sigma_{i}^{k}$ that dominate $\sigma_{i}$ relative to $\Sigma_{-i}^{m}$. Suppose $\dom_{i}^{0} \neq \emptyset$. We will show that, for each $k=0,\ldots,m$, $\dom_{i}^{k} \neq \emptyset$. 

The case of $k=0$ is immediate. Assume the claim holds for $k < m$, i.e., $\dom_{i}^{k} \neq \emptyset$. Define the function $v_{i}^{k} : \Sigma_{i}^{k} \rightarrow  \mathbb{R}$ so that 
\[
	v_{i}^{k}(\rho_{i}) = \min_{\sigma_{-i} \in \Sigma_{-i}^{m}} \Upsilon_{i}^{m}(\sigma_{i},\rho_{i},\sigma_{-i}).
\]
Since $\Sigma_{-i}^{m}$ is compact (Lemma \ref{lem:m-closed}) and $\Upsilon_{i}^{m}$ is continuous, it follows from Berge's Theorem that $v_{i}^{k}$ is continuous. Moreover, since $\dom_{i}^{k} \neq \emptyset$, there exists some $\rho_{i}$ with $v_{i}^{k}(\rho_{i}) > 0$.

Since $\Sigma_{i}^{m}$ is compact (Lemma \ref{lem:m-closed}) and $v_{i}^{k}$ is continuous, there exists  $\rho_{i}^{*} \in \Sigma_{i}^{k}$ so that 
\[
	 v_{i}^{k}(\rho_{i}^{*}) \geq  v_{i}^{k}(\rho_{i})
\]
for all $\rho_{i} \in \Sigma_{i}^{k}$. As a consequence, $v_{i}^{k}(\rho_{i}^{*})>0$, i.e., 
\[
	u_{i}(\rho_{i}^{*},\sigma_{-i}) > u_{i}(\sigma_{i},\sigma_{-i}) \qquad \text{ for all } \sigma_{-i} \in \Sigma_{-i}^{m}.
\]

It suffices to show that $\rho_{i}^{*} \in  \Sigma_{i}^{k+1}$. Suppose not, i.e., there exists $\mu_{i} \in \Sigma_{i}^{k}$ so that 
\[
	u_{i}(\mu_{i},\sigma_{-i}) > u_{i}(\rho_{i}^{*},\sigma_{-i}) \qquad \text{ for all } \sigma_{-i} \in \Sigma_{-i}^{k}.
\]
Since $\Sigma_{-i}^{m} \subseteq \Sigma_{-i}^{k}$,  
\[
	u_{i}(\mu_{i},\sigma_{-i}) > u_{i}(\rho_{i}^{*},\sigma_{-i}) > u_{i}(\sigma_{i},\sigma_{-i}) \qquad  \text{ for all } \sigma_{-i} \in \Sigma_{-i}^{m}.
\]
or
\[
	\Upsilon_{i}^{m}(\sigma_{i},\mu_{i},\sigma_{-i}) >	\Upsilon_{i}^{m}(\sigma_{i},\rho_{i}^{*},\sigma_{-i}) \qquad  \text{ for all } \sigma_{-i} \in \Sigma_{-i}^{m}.
\]
Since $\Sigma_{-i}^{m}$ is compact (Lemma \ref{lem:m-closed}) and $\Upsilon_{i}^{m}$ is continuous, there exists some $\varepsilon >0 $ so that 
\[
	\Upsilon_{i}^{m}(\sigma_{i},\mu_{i},\sigma_{-i}) \geq	\Upsilon_{i}^{m}(\sigma_{i},\rho_{i}^{*},\sigma_{-i}) + \varepsilon \qquad  \text{ for all } \sigma_{-i} \in \Sigma_{-i}^{m}.
\]
But, then 
\[
	v_{i}^{k}(\mu_{i}) \geq v_{i}^{k}(\rho_{i}^{*}) + \varepsilon > v_{i}^{k}(\rho_{i}^{*}),
\]
contradicting the fact that $\rho_{i}^{*}$ maximizes $v_{i}^{k}$. 
\end{proof}

\end{document}